\documentclass[runningheads]{llncs}
\usepackage[T1]{fontenc}
\usepackage{graphicx}
\usepackage{amsmath}
\usepackage{amssymb}
\usepackage{gensymb}
\DeclareMathOperator*{\argmax}{arg \; max} 
\usepackage[hidelinks]{hyperref}

\begin{document}
\title{Resolving quantitative MRI model degeneracy in self-supervised machine learning}
\titlerunning{Resolving qMRI model degeneracy in self-supervised ML}
%
\author{Giulio V. Minore\inst{1,2}\orcidID{0009-0001-4529-8547}\thanks{Corresponding author: giulio.minore.17@ucl.ac.uk}
\and
Louis Dwyer-Hemmings\inst{1,3,4}\orcidID{0000-0001-8449-2660}
\and
Timothy J.P. Bray\inst{1,3,4}\orcidID{0000-0001-8886-5356}
\and
Hui Zhang\inst{1,5}\orcidID{0000-0002-5426-2140}
}

\authorrunning{G.V. Minore et al.}
%
\institute{Hawkes Institute, University College London, London, United Kingdom
\and
Department of Medical Physics and Biomedical Engineering, University College London, London, United Kingdom
\and
Centre for Medical Imaging, University College London, London, United Kingdom
\and
Department of Imaging, University College London Hospital, London, United Kingdom
\and
Department of Computer Science, University College London, London, United Kingdom
}
\maketitle              
\begin{abstract}
Quantitative MRI (qMRI) estimates tissue properties of interest from measured MRI signals. This process is conventionally achieved by model fitting, whose computational expense limits qMRI’s clinical use, motivating recent development of machine learning-based methods. Self-supervised approaches are particularly popular as they avoid the pitfall of distributional shift that affects supervised methods. However, it is unknown how such methods behave if similar signals can result from multiple tissue properties, a common challenge known as model degeneracy. Understanding this is crucial for ascertaining the scope within which self-supervised approaches may be applied. To this end, this work makes two contributions. First, we demonstrate that model degeneracy compromises self-supervised approaches, motivating the development of mitigation strategies. Second, we propose a mitigation strategy based on applying appropriate constraining transforms on the output of the bottleneck layer of the autoencoder network typically employed in self-supervised approaches. We illustrate both contributions using the estimation of proton density fat fraction and $R_2^*$ from chemical shift-encoded MRI, an ideal exemplar due to its exhibition of degeneracy across the full parameter space. The results from both simulation and \textit{in vivo} experiments demonstrate that the proposed strategy helps resolve model degeneracy.

\end{abstract}

\section{Introduction}

Quantitative magnetic resonance imaging (qMRI) aims to improve on conventional MRI, which primarily relies on image contrast due to its signal being an aggregation of tissue properties. By disentangling these contributions to the signal, qMRI provides quantitative, more disease-specific measures of the underlying tissue properties, which increases the interpretability of scans, facilitating diagnosis and assessment of disease.

Conventionally, qMRI parameter estimation is performed through voxel-wise model fitting of the measured MRI signal. However, this is computationally expensive as it relies on iteratively and independently maximising some likelihood function for each voxel, reducing its clinical utility.

To improve on the computational requirements of conventional parameter estimation, machine learning, and especially deep learning-based methods have been proposed. Deep learning approaches generally employ supervised strategies~\cite{Golkov,Bertleff,Grussu,Gyori,RAIDER}, relying on ground truth parameter labels. However, their performance may be affected by distributional shift between the training and target measured signal. This could be caused by discrepancies between the distributions of training and target parameter combinations~\cite{Gyori,Epstein_2024} or signal-to-noise ratio (SNR). To avoid any potential distributional shift, self-supervised methods have been proposed~\cite{Barbieri,Kaandorp_Barbieri,Grussu,ssVERDICT}. These do not require ground truth labels and can be used to train a network directly on the target data.

Despite these theoretical advantages, to the best of our knowledge, there have been no studies that have investigated how self-supervised methods would behave if similar signals can result from multiple tissue properties, a common challenge known as model degeneracy. This limits our understanding of the scope within which such strategies may be used. In this work, we first describe the impact of model degeneracy on self-supervised algorithms, then propose a method to mitigate its effect. These contributions are demonstrated both \textit{in silico} and \textit{in vivo} chemical shift-encoded MRI (CSE-MRI) data, which will serve as an exemplar of model degeneracy due to its presence throughout the parameter space.

\section{Theory}

In this section, following an introduction to the qMRI parameter estimation problem, we describe the effect of model degeneracy on self-supervised algorithms and the proposed method to mitigate this problem.

\subsection{qMRI parameter estimation problem}

The aim of qMRI parameter estimation is to infer the unknown tissue properties $\boldsymbol{y} \in \mathbb{R}^{N_y}$ at a given voxel from the measured signal $\boldsymbol{S}\in \mathbb{R}^{N_S}$ at that voxel acquired with the settings $\boldsymbol{Z}$, where $N_y$ is the number of distinct tissue properties and $N_S$ is the number of signal measurements. 

This problem has been typically solved by conventional fitting. This relies on fitting a biophysical forward model $\mathcal{M}$ that can predict the noise-free signal corresponding to some tissue properties $\boldsymbol{\Tilde{y}}$, such that $\boldsymbol{\Tilde{S}} = \mathcal{M}(\boldsymbol{Z}; \boldsymbol{\Tilde{y}})$. The parameter estimates $\boldsymbol{\hat{y}}$ can then be found by maximising some appropriate likelihood function such that
\begin{equation}
    \boldsymbol{\hat{y}} = \argmax_{\boldsymbol{\Tilde{y}}} \; \mathcal{L}(\mathcal{M}(\boldsymbol{Z}; \boldsymbol{\Tilde{y}})\vert\boldsymbol{S} ).
\end{equation}
In the presence of model degeneracy, the above likelihood will possess multiple local maxima, which must be resolved by fitting from multiple initial guesses to the parameter estimates.

\subsection{Self-supervised method and the impact of model degeneracy}

Self-supervised algorithms have been recently proposed to accelerate qMRI parameter estimation~\cite{Barbieri,Kaandorp_Barbieri,Grussu,ssVERDICT}. Similar to the more common supervised alternatives, they aim to find a function $f_{\boldsymbol{\theta}}: \mathbb{R}^{N_S}\rightarrow\mathbb{R}^{N_y}$ parameterised by $\boldsymbol{\theta}$ (e.g., neural network weights) that can directly map a signal measured at a voxel to some tissue properties that are as close as possible to the underlying tissue properties at the voxel. However, unlike supervised alternatives, self-supervised algorithms use the predicted tissue properties $\boldsymbol{\Tilde{y}} = f_{\boldsymbol{\theta}}(\boldsymbol{S})$ to reconstruct the noise-free signal using the forward model, such that $\boldsymbol{\Tilde{S}} = \mathcal{M}(\boldsymbol{Z};\boldsymbol{\Tilde{y}})$.

\begin{figure}
\includegraphics[width=\textwidth]{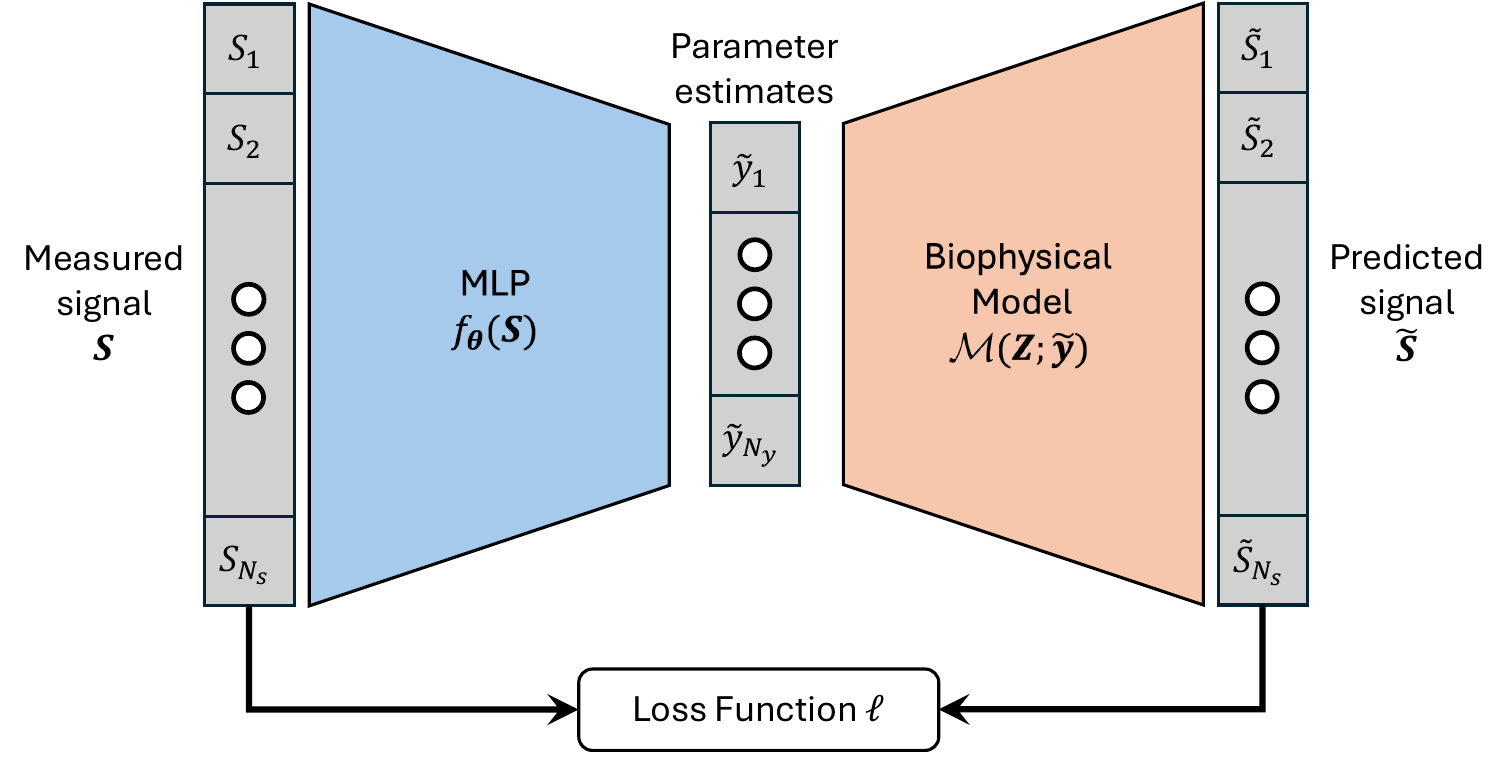}
\caption{Diagram of a physics-based autoencoder. A measured signal $\boldsymbol{S}$ is fed to a neural network which predicts tissue properties $\boldsymbol{\Tilde{y}}$. These are then used to reconstruct the signal $\boldsymbol{\Tilde{S}}$ using the forward model, which is used to calculate the self-supervised training loss.}\label{fig:SSL_sol}
\end{figure}

This approach has been commonly considered as being based on a physics-based autoencoder (see Fig.~\ref{fig:SSL_sol}). Similar to standard autoencoders, a neural network implements the encoder to map some input signal to the output of the bottleneck layer representing the parameter estimates $\boldsymbol{\Tilde{y}}$. Unlike standard autoencoders, the decoder is not implemented with another neural network but with the forward model which as a result does not possess any learnable parameters. The loss is defined as a negative-likelihood between $N$ measured signals $\boldsymbol{S}_i$ and predicted signals $\boldsymbol{\Tilde{S}}_i= \mathcal{M}(\boldsymbol{Z};f_{\boldsymbol{\theta}}(\boldsymbol{S}_i))$, such that
\begin{equation}
            \ell=-\frac{1}{N}\sum_{i=1}^N \mathcal{L}( \boldsymbol{\Tilde{S}}_i \vert \boldsymbol{S}_i)
            =-\frac{1}{N}\sum_{i=1}^N \mathcal{L}( \mathcal{M}(\boldsymbol{Z};f_{\boldsymbol{\theta}}(\boldsymbol{S}_i)) \vert \boldsymbol{S}_i).
    \label{eq:loss_SSL}
\end{equation}

While the effect of model degeneracy has been described for supervised methods~\cite{Bishop,Guerreri}, this has not yet been done for self-supervised algorithms. To help understand this effect, similar to~\cite{Guerreri}, we begin with a simplifying theoretical analysis where we assume the strong form of model degeneracy, namely that there exists at least one subset $\boldsymbol{Y_d}$ in which different tissue properties $\boldsymbol{y}$ produce the same signal $\boldsymbol{S}_d$, such that
\begin{equation}
    \forall \; \boldsymbol{y}_j,\boldsymbol{y}_k \in \boldsymbol{Y_d}, \; \boldsymbol{y}_j \neq \boldsymbol{y}_k \; \text{and}\; \boldsymbol{S}_j = \boldsymbol{S}_k = \boldsymbol{S}_d .
\end{equation}
Without loss of generality, assuming the first $n$ samples belong to the degenerate subset, the loss function may be re-written as
\begin{equation}
    \begin{split}
            \ell&=-\frac{1}{N}\sum_{i=1}^n \mathcal{L}( \mathcal{M}(\boldsymbol{Z};f_{\boldsymbol{\theta}}(\boldsymbol{S}_i)) \vert \boldsymbol{S}_i)-\frac{1}{N}\sum_{i=n+1}^N \mathcal{L}( \mathcal{M}(\boldsymbol{Z};f_{\boldsymbol{\theta}}(\boldsymbol{S}_i)) \vert \boldsymbol{S}_i).
    \end{split}
    \label{eq:loss_SSL_degen}
\end{equation}
By construction, for $1\leq i \leq n$, $\boldsymbol{S}_i = \boldsymbol{S}_d$. Since both the encoder (a neural network) and the decoder (a forward model) are functions and therefore cannot represent one-to-many mappings, the likelihood terms for the first $n$ samples are identical. 

Assuming the two terms in the loss can be independently minimised, any unique combination of tissue properties $\boldsymbol{y}$ in $\boldsymbol{Y_d}$ produces the signal that minimises the first term. Therefore, the network can learn to predict any individual combination in $\boldsymbol{Y_d}$ at random, neglecting all the other possible solutions. Given this result for the strong form of degeneracy, we hypothesise that a similar behaviour will manifest in the presence of the general form of degeneracy.

\subsection{Proposed method to resolve degeneracy}
\label{proposed_method}

The proposed method is inspired by previous approaches developed for supervised algorithms~\cite{Bishop,RAIDER} but with a novel adaptation to support the unique demand of self-supervised algorithms. 

The existing approaches for supervised algorithms work by assuming that the parameter space can be subdivided such that within each sub-space degeneracy is no longer present. In~\cite{Bishop}, the parameter space is split into two; after rejecting one of the sub-spaces as containing only non-feasible parameter values, a neural network is trained with training labels drawn solely from the other sub-space. Later in~\cite{RAIDER}, the parameter space is also split into two; but because both sub-spaces contain feasible parameter values, two neural networks are trained such that each sub-space has its own designated network trained with training labels drawn solely from the sub-space itself.
At inference time, the outputs from the two networks are compared in terms of their likelihoods to choose the best parameter estimate. The present work builds on this general multi-network approach.

However, this approach cannot be directly applied to self-supervised methods. This is because training data for self-supervised and supervised methods are fundamentally different. For supervised methods, the training data consist of feature-label pairs, with the (input) feature being the signal and the (output) label being tissue properties.  This allows the training data to be separated into distinct subsets with each containing labels belonging to some chosen subset of parameters. In contrast, the training data for self-supervised methods consist of feature-label pairs with both the (input) feature and the (output) label being the signal. As a result, the existing approaches for supervised methods are no longer applicable.

As an alternative, we propose to add a constraining transform to the bottleneck layer. While such transforms are commonly used in conventional fitting to ensure the physical plausibility of parameter estimates, here we take advantage of them to guarantee network predictions belong to a sub-space that does not exhibit degeneracy. To understand how this works, recall that for the existing self-supervised methods, the output of the bottleneck layer represents the predicted tissue properties such that $\boldsymbol{\Tilde{y}} = f_{\boldsymbol{\theta}}(\boldsymbol{S})$. With the introduction of the constraining transform $\mathcal{C}$, the output following this transform now represents the predicted tissue properties such that $\boldsymbol{\Tilde{y}} = \mathcal{C}(f_{\boldsymbol{\theta}}(\boldsymbol{S}))$. 

\section{Methods}

This section describes an implementation of the proposed multi-network method for CSE-MRI data to demonstrate that model degeneracy may be resolved in self-supervised learning. This is performed both in simulation and \textit{in vivo}, and a description of both datasets and experiments will be given.

\subsection{An example of qMRI model degeneracy: CSE-MRI}

We now illustrate model degeneracy in CSE-MRI, which will serve as an exemplar to evaluate our proposed method to help resolve degeneracy. CSE-MRI is used to separate the contributions of water and fat to the signal in a single voxel, denoted $\rho_W$ and $\rho_F$, typically from multi-echo gradient echo acquisition data. It aims to estimate proton density fat fraction $(\text{PDFF}=\frac{\rho_F}{\rho_W+\rho_F})$ and the $R_2^*$ relaxation rate, which are useful biomarkers across a wide range of diseases and tissues~\cite{Starekova_liver_fat,YoonJeongHee_pancreas,Bray_inflamed_bone,Latifoltojar_myeloma,Morrow_neuromuscular,Reeder_iron_overload}. Although their estimation has traditionally required complex data~\cite{ReederIdealGRE,Yu_T2star_IDEAL,Hernando_graph_cut}, magnitude-fitting techniques~\cite{MAGO,MAGORINO} have gained interest due to the phase data sometimes being unreliable or unavailable.

The forward model for the magnitude signal from CSE-MRI is given by
\begin{equation}
    \mathcal{M}(t \vert S_0, \text{PDFF}, R_2^*) = S_0\Big\vert (1-\text{PDFF})+\text{PDFF} \sum_{k=1}^K r_k e^{(i 2\pi f_{F,k} t)}\Big\vert e^{(-t R_2^*)}
\label{eq:forward_model}
\end{equation}
where $S_0$ is the signal at time $t=0$, $K$ is the number of spectral fat components, $r_k$ and $f_{F,k}$ are respectively the relative amplitude and frequency shift of each spectral fat component and are defined \textit{a priori}.

Assuming the MRI magnitude signal is Rician-distributed, the likelihood of a predicted signal $\boldsymbol{\Tilde{S}}_i =  \mathcal{M}(t \vert \boldsymbol{\Tilde{y}}_i)$ in voxel $i$, reconstructed from predictions $\boldsymbol{\Tilde{y}}_i$, given a measured signal $\boldsymbol{S}_i$ is~\cite{Sijbers1998}
\begin{equation}
    \mathcal{L}_{\text{Rice}}(\boldsymbol{\Tilde{S}}_i, \sigma^2 \vert \boldsymbol{S}_i)=\prod_{k=1}^{N_{TE}} \frac{S_{ik}}{\sigma^2}  \exp\Big(\frac{-(S_{ik}^{2}+\Tilde{S}_{ik}^2)}{2\sigma^2}\Big)I_0\Bigl(\frac{S_{ik} \Tilde{S}_{ik}}{\sigma^2}\Bigl)
\end{equation}
where $N_{TE}$ is the number of echo times at which the signal is acquired, $\sigma$ is the standard deviation of the underlying complex Gaussian noise and $I_0$ is the modified Bessel function of the first kind with order zero.

\begin{figure}
\includegraphics[width=\textwidth]{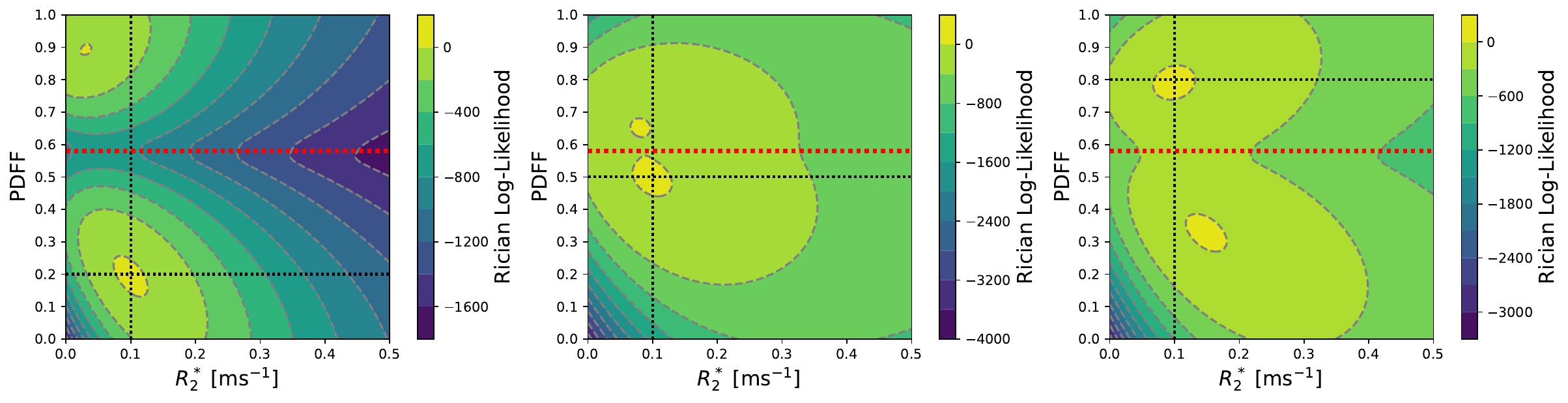}
\caption{Rician log-likelihood as a function of PDFF and $R_2^*$ for different tissue parameter combinations. Ground truths are indicated as dotted black lines, representing $R_2^* = 0.1$ ms$^{-1}$ and PDFF equal to 0.2 (left), 0.5 (centre) 0.8 (right). There are two maxima in likelihood, occurring above or below the red dotted switching line at PDFF = 0.58, giving rise to degeneracy.
} \label{fig:rician_log_likelihood}
\end{figure}

This forward model is known to exhibit systematic degeneracy with the presence of two local maxima in the likelihood function. These two maxima occur for either a low (closer to 0) or a high (closer to 1) PDFF, approximately opposite on the PDFF spectrum (see Fig.~\ref{fig:rician_log_likelihood}). These will henceforth be referred to as the water-dominant or fat-dominant solutions. Importantly, it has been shown that the PDFF space can be divided into two: a water-dominant space ($\text{PDFF} < 0.58$) and a fat-dominant one ($\text{PDFF} \geq  0.58$), such that each degenerate pair is split into either the water-dominant space or the fat-dominant one~\cite{MAGO,MAGORINO,RAIDER}. As a result, within each of these two spaces, degeneracy is no longer present.

While model degeneracy has been resolved in conventional fitting and supervised learning methods for magnitude CSE-MRI~\cite{MAGO,MAGORINO,RAIDER}, to the best of our knowledge, no self-supervised technique has been developed for this problem. Here we apply the proposed method described in Sec~\ref{proposed_method} to implement a dual network approach. For each network, we introduce an appropriate constraining transform for PDFF, such that each network respectively makes predictions in the water- or fat-dominant space. The constraining transforms are
\begin{equation}
\mathcal{C}(x) =
    \begin{cases}
        \frac{b}{2} (\sin(x) + 1)  & \;\text{for the water-dominant network}
        \\
        1 - \frac{(1-b)}{2}(\sin(x)+ 1)  &\;        \text{for the fat-dominant network}
    \end{cases}
    \label{eq:pdff_constraint}
\end{equation}
where $b = 0.58$ corresponds to the boundary between the degenerate sub-spaces. For the water-dominant network (the `water' network in short), its PDFF estimates are constrained to lie in the [0,b] interval, and for the fat-dominant network (the `fat' network in short), its PDFF estimates are constrained to lie in the [b,1] interval.
The $S_0$ and $R_2^*$ parameters are also constrained to be non-negative by the squaring transform, to ensure that the predictions are physically plausible.

\begin{figure}
\includegraphics[width=\textwidth]{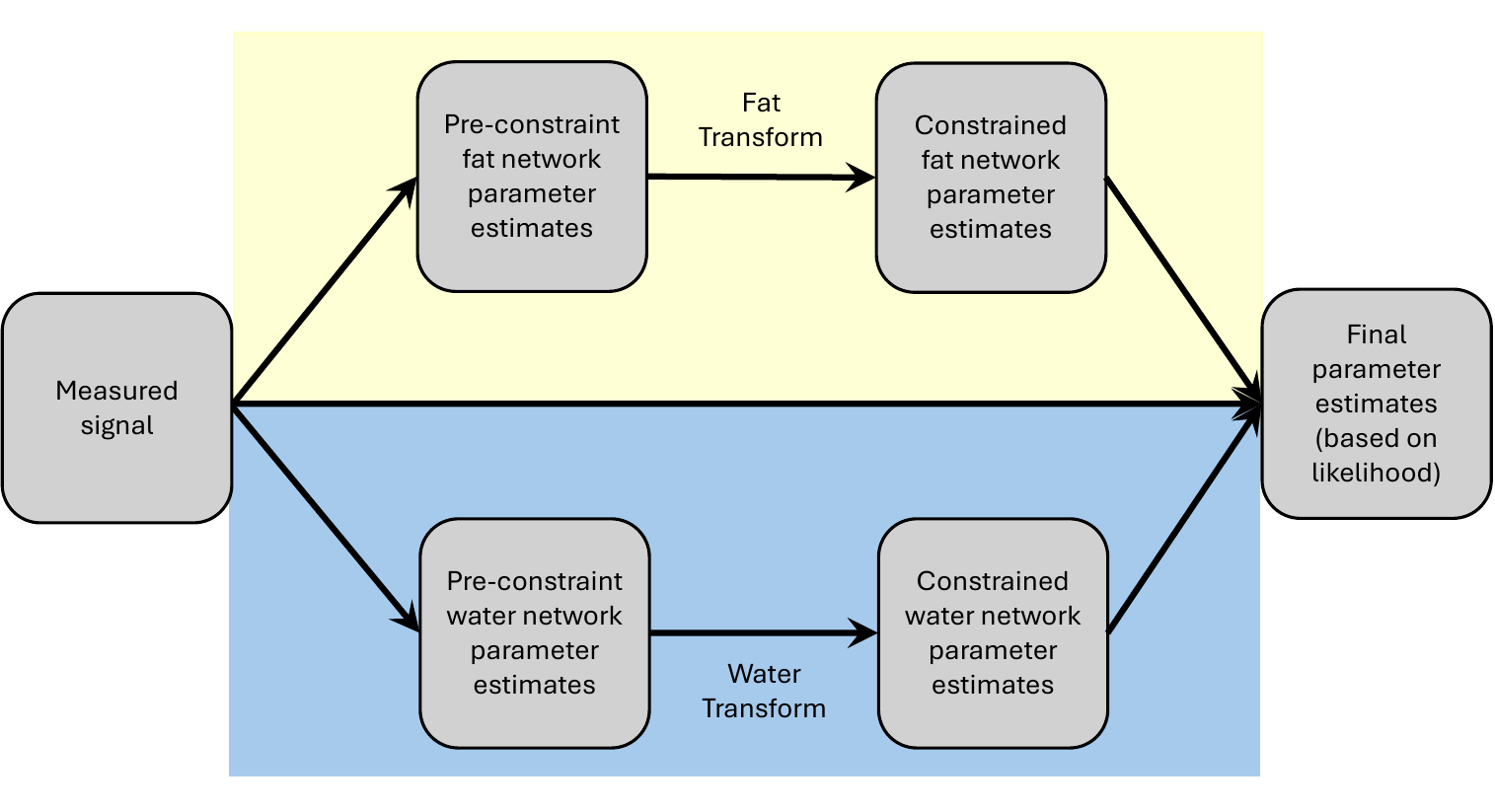}
\caption{Inference procedure using dual network approach. The measured signal is fed to the `water' and `fat' networks that make predictions in their respective parameter sub-space via the use of constraining transforms. The final parameter estimates are selected by comparing their likelihood.
} \label{fig:inference}
\end{figure}

As in~\cite{RAIDER}, the final parameter estimates are selected based on the candidate solution maximising the likelihood (see Fig.~\ref{fig:inference}).

\subsection{Implementation}

\subsubsection{Architecture}
The self-supervised encoder network architecture was chosen to resemble previous self-supervised work~\cite{Barbieri,ssVERDICT}. A multilayer perceptron takes a voxel-wise signal as input, which is then fed through five fully-connected layers, using exponential linear unit activation functions~\cite{ELU}. All but the last layer possess $N_{TE}$ nodes, corresponding to the number of measurements, as was previously done in other self-supervised approaches. The final layer has three output nodes, corresponding to the three parameters of interest (namely $S_0$, PDFF and $R_2^*$).

\subsubsection{Training}

Self-supervised networks are directly trained using the target data on which predictions are made, both in \textit{silico} and \textit{in vivo}. This ensures that training and target data are identical, such that distributional shift cannot occur. Networks are optimised by minimising the negative Rician log-likelihood loss~\cite{Parker_rician_loss} due to Rician distribution of magnitude signals. Training is performed using the Adam optimizer~\cite{Adam} with learning rate 0.001, minibatch size of 128, patience of 10 as in~\cite{Barbieri}. The code is available on GitHub\footnote{\url{https://github.com/CIG-UCL/RAIDER-SSL}}.

\subsection{Experiments}

\subsubsection{\textit{In silico}}
We first investigate the effects of model degeneracy in synthetic data, where ground truth parameter values are known. To this end, we simulate signals with the forward model in Eq.~(\ref{eq:forward_model}), using the acquisition settings corresponding to the \textit{in vivo} dataset described below. This synthetic data is generated by producing different combinations of PDFF and $R_2^*$, evenly sampling 100 PDFF values in the [0,1] interval and 50 $R_2^*$ values in the [0,0.5] ms$^{-1}$ interval. For each of these 100$\times$50 combinations, 100 noisy signals are simulated by adding Gaussian noise to the underlying complex signal before taking its magnitude, such that the ground truth SNR is 60 (typical for CSE-MRI acquisitions~\cite{MAGO,MAGORINO,RAIDER}).

We train single networks without constraining the PDFF predictions to lie within a specific sub-space (i.e. PDFF $\in [0,1]$) on synthetic data, where ground-truth tissue properties are known. We hypothesise that this naive implementation will unpredictably learn to output either the water-dominant or fat-dominant solutions. To evaluate the proposed method designed to ensure both the `water' and `fat' networks are learned, we also train two networks using the constraining transforms in Eq.~(\ref{eq:pdff_constraint}).

Finally, the parameter estimation performance of the proposed method is evaluated in terms of accuracy (bias) and precision (standard deviation) on the synthetic dataset.

\subsubsection{\textit{In vivo}}

The \textit{in vivo} dataset consists of the imaged pelvis of a healthy volunteer who gave informed consent. The data was acquired on a 3T Philips Ingenia scanner using a multi-echo 3D spoiled gradient echo sequence with 6 echoes ($TE_1 = 1.15$ ms, $\Delta_{TE} = 1.15$ ms, $TR = 8.5$ ms, flip angle = 3$\degree$). 50 slices of thickness 1.5 mm were acquired, with matrix size 240 $\times$ 240 and voxel size 1.8 mm $\times$ 1.8 mm. To create pseudo-ground truths for the tissue properties \textit{in vivo}, the data was supersampled by repeating the acquisition 4 times within a single scanning session, with each acquisition using two $k$-space averages to increase SNR. Conventional qMRI parameter estimation using the MAGORINO algorithm~\cite{MAGORINO} was then performed on this supersampled dataset to provide pseudo-ground truths. In the same scanning session, the subject was also imaged using a single $k$-space average. This is representative of a clinical acquisition and serves as evaluation data to compare the parameter maps predicted by different qMRI methods to the pseudo-ground truths.

Parameter estimation is performed on the single $k$-space average \textit{in vivo} image and compared to the pseudo-ground truths. A manually-selected threshold is used to remove background voxels, which are not of interest and may affect the self-supervised procedure. We first use a naive self-supervised implementation using a single network to show that the network learns to predict tissue properties belonging to only one sub-space. The dual network approach is then implemented to help resolve model degeneracy. Supervised-based~\cite{RAIDER} and conventional fitting~\cite{MAGORINO} are also performed, and the resulting PDFF, $R_2^*$ and $S_0$ maps from these three methods are compared to the pseudo-ground truths.

\section{Results}

\begin{figure}[h!]
\includegraphics[width=\textwidth]{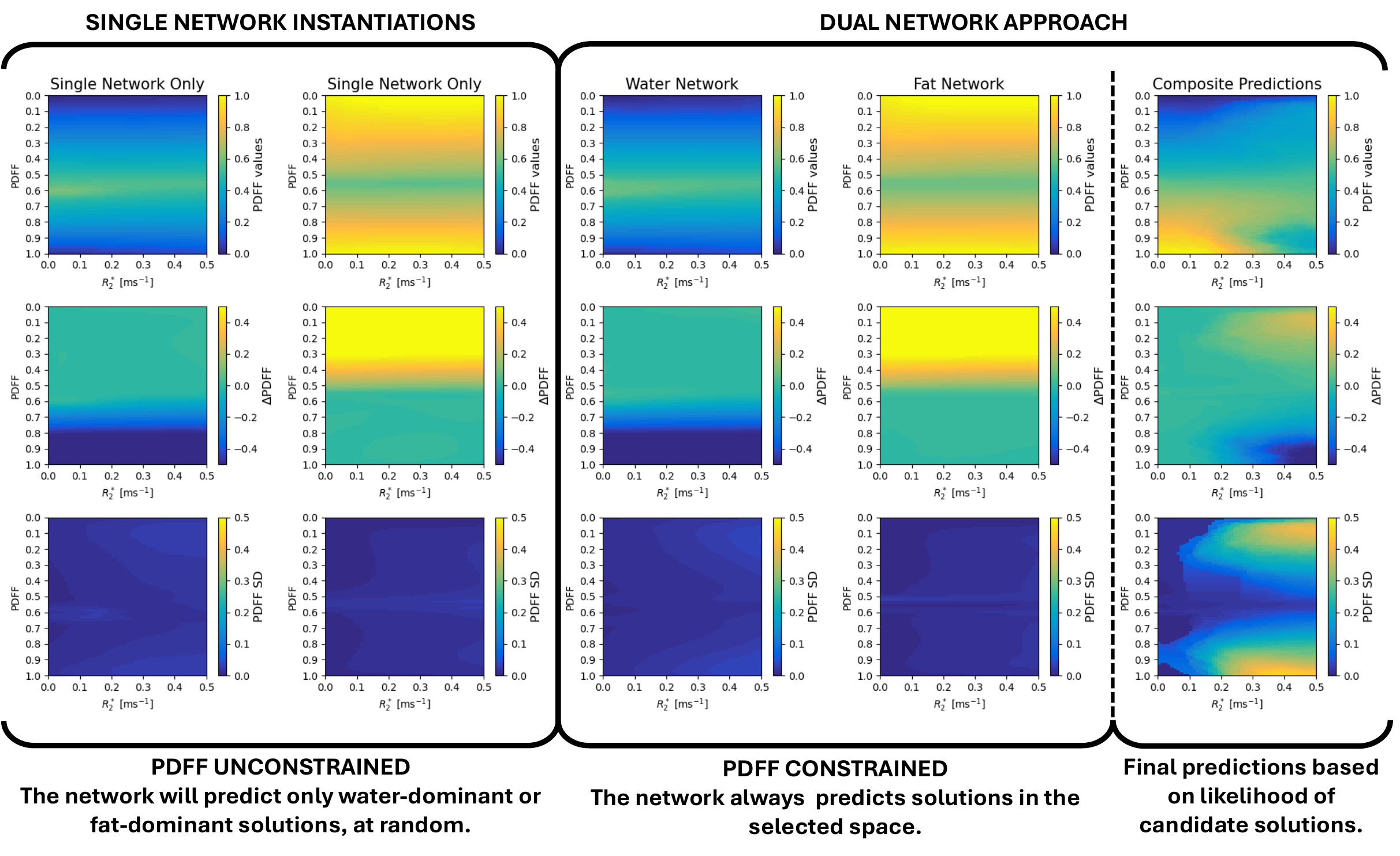}
\caption{Feasibility of self-supervised learning in the presence of model degeneracy \textit{in silico}. Without the proposed constraining transforms, self-supervised networks can unpredictably learn to output either water- ($1^{\text{st}}$ column) or fat-dominant ($2^{\text{nd}}$ column) solutions. With the constraining transforms, the target networks are predictably learned ($3^{\text{th}}$ and $4^{\text{th}}$ columns). The prediction that maximises likelihood is chosen as the estimate ($5^{\text{th}}$ column). Rows sequentially represent predicted PDFF values, bias and standard deviation.
} \label{fig:pdff_comp}
\end{figure}

Fig.~\ref{fig:pdff_comp} shows the \textit{in silico} results of PDFF estimation for different self-supervised learning strategies. Using a single network approach, without constraining PDFF to lie within one of the two degenerate sub-spaces, results in the trained network to unpredictably make either water- or fat-dominant predictions. This phenomenon occurred regardless of whether a sub-space was rejected from the training data. That is, the network can learn to predict either solution when trained on data from the full, water-dominant or fat-dominant parameter spaces. However, training networks using the proposed constraining transforms ensures the networks make predictions in the selected parameter sub-space.

\begin{figure}[h!]
\includegraphics[width=\textwidth]{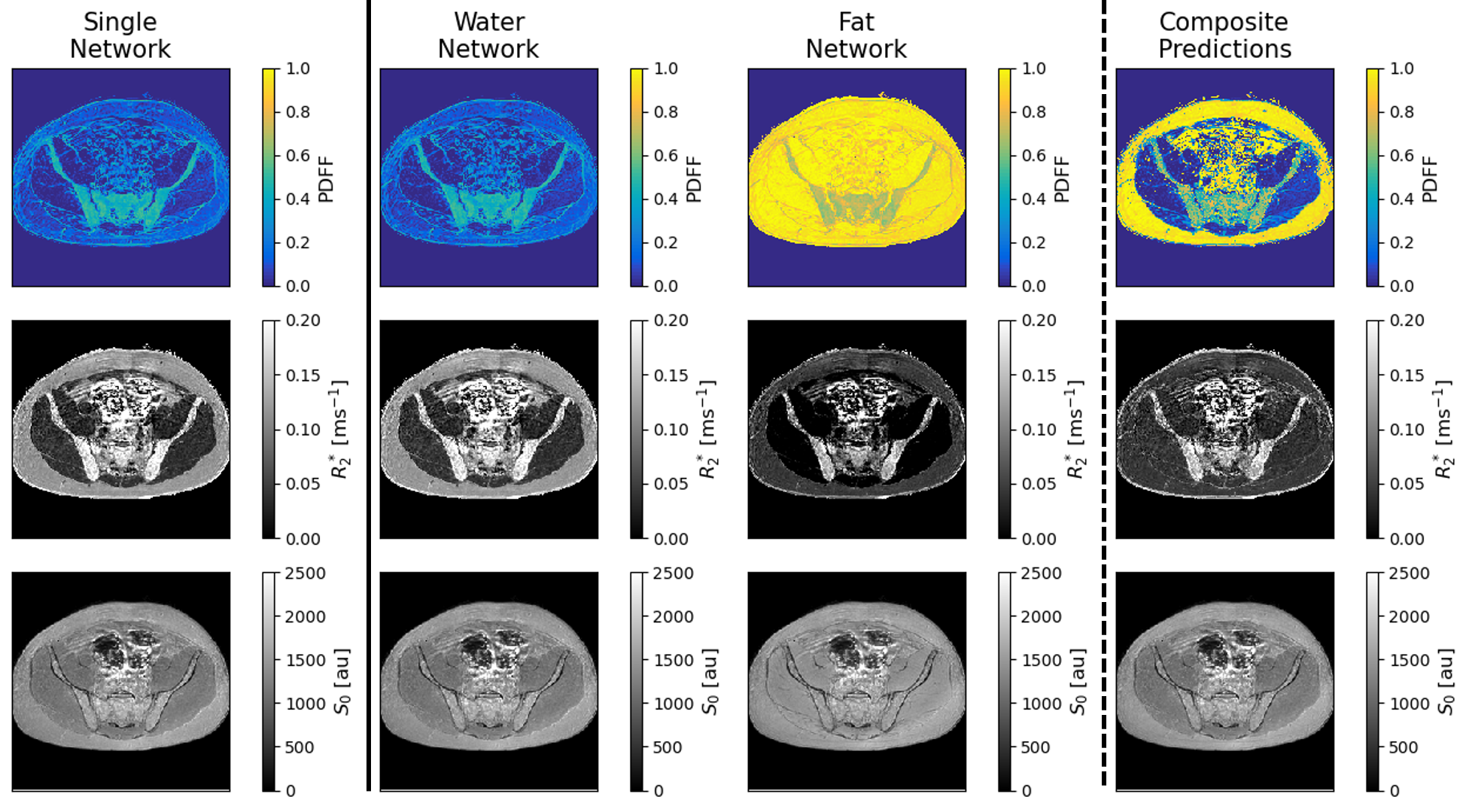}
\caption{Degeneracy mitigation \textit{in vivo} using a dual network approach. A single network (leftmost) unpredictably output tissue properties in one degenerate sub-space. The proposed use of constraining transforms ensures `water' and `fat' networks are learned with certainty (middle two). The prediction that maximises likelihood is chosen as the estimate (rightmost).
} \label{fig:SSL_in_vivo}
\end{figure}

Fig.~\ref{fig:SSL_in_vivo} shows that using a single self-supervised network does not resolve degeneracy \textit{in vivo}, since only water-dominant solutions are predicted. By training constrained `water' and `fat' networks, the candidate solutions can be used to create composite predictions based on likelihood.


\begin{figure}
\centering
\includegraphics[width=0.8\textwidth]{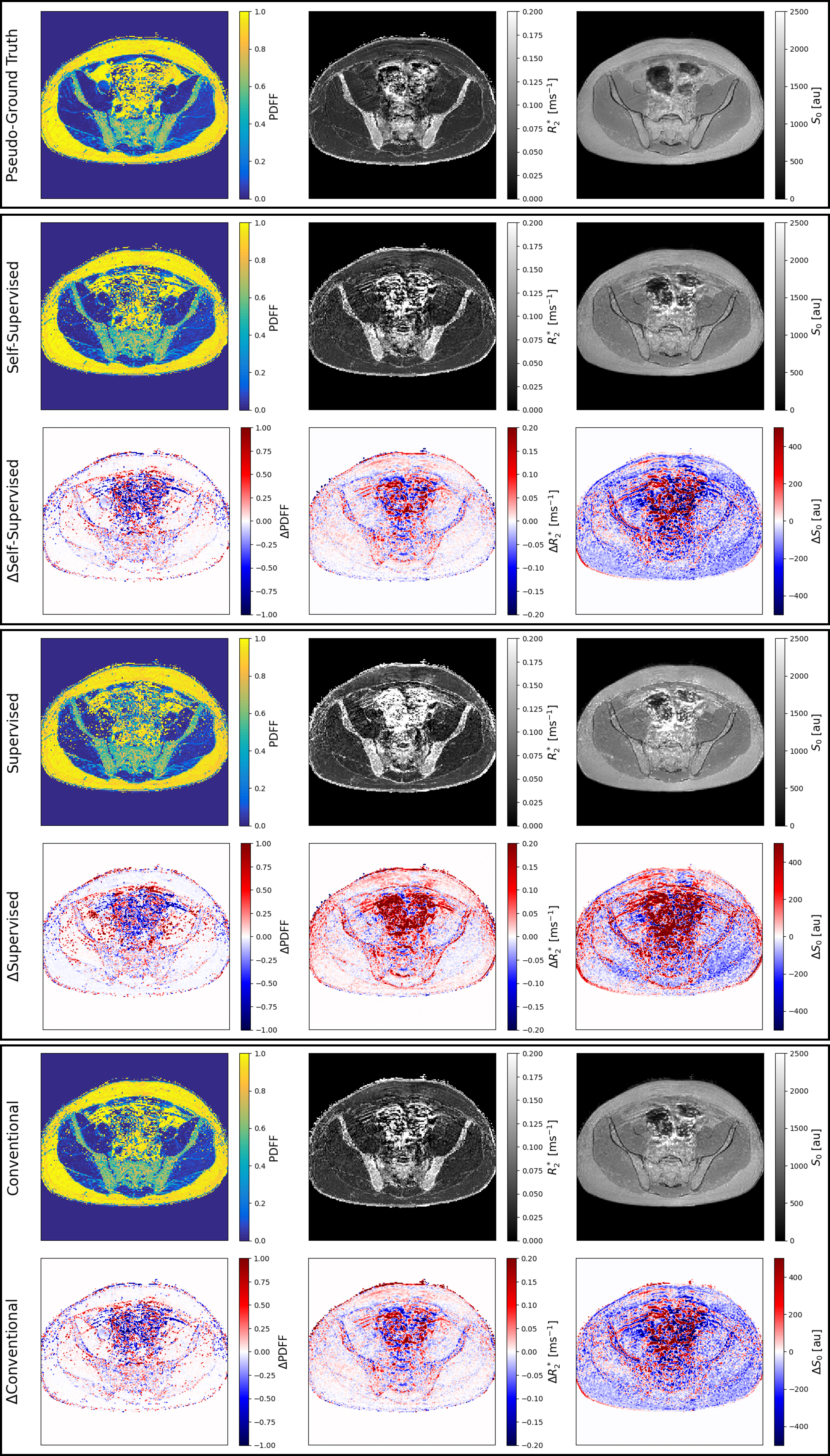}
\caption{Estimated parameter maps using self-supervised ($2^{\text{nd}}$ row), supervised ($4^{\text{th}}$ row) and conventional ($6^{\text{th}}$ row) methods. Their difference with the pseudo-ground truths obtained from the supersampled dataset ($1^{\text{st}}$ row) are also shown.
} \label{fig:in_vivo_comp}
\end{figure}

Fig.~\ref{fig:in_vivo_comp} shows the pseudo ground-truths acquired from the supersampled dataset, as well as the parameter maps estimated using self-supervised, supervised and conventional methods. Overall, the self-supervised method has lower bias than its supervised alternative for the clinically relevant PDFF and $R_2^*$, suggesting self-supervised strategies may be preferred over supervised ones, as they avoid distributional shift that can negatively affect parameter estimation. Indeed, self-supervised achieves lower PDFF bias in comparison to supervised, particularly visible in subcutaneous fat. In the same region, the supervised method exhibits larger $R_2^*$ bias, while $S_0$ predictions are also larger, likely due to compensation for the higher predicted $R_2^*$ decay. Furthermore, self-supervised and conventional techniques achieve similar performance across different tissue types. Although all three methods show large bias in the bowel, this is expected due to peristalsis (movement of the intestine due to involuntary muscle contractions).

\section{Discussion \& Conclusions}

In this work, we make two contributions to the understanding of the scope within which self-supervised methods may be applied for qMRI parameter estimation. First, we describe the behaviour of self-supervised algorithms in the presence of model degeneracy and confirm it in CSE-MRI data, both in simulation and \textit{in vivo}. Neural networks trained on degenerate signals unpredictably learn to output a unique combination of tissue properties that produces that signal, compromising parameter estimation.

Second, we propose a method to help resolve degeneracy for self-supervised approaches by taking inspiration from previously developed strategies for supervised algorithms~\cite{Bishop,RAIDER}. Both these strategies rely on the assumption that the parameter space can be split into sub-spaces, such that no degeneracy exists within each sub-space. Networks designated to a sub-space can then be trained by sampling labels from the sub-space itself to avoid the pitfalls of model degeneracy. However, these strategies cannot be directly applied to self-supervised methods due to the fundamental difference between the training data of these different machine learning strategies. On one hand, separating labels from degenerate sub-spaces may be effective for supervised learning since it relies on feature-label pairs of (input) features being the signal and (output) labels being tissue properties. On the other hand, both the features and labels of self-supervised algorithms are the signal, making this strategy inapplicable. Instead, we use a constraining transform to the bottleneck of the autoencoder, such that the estimated tissue properties lie in one of these sub-spaces. Using these constraining transforms and a dual network approach inspired by~\cite{RAIDER}, we demonstrate that degeneracy can be mitigated using self-supervised methods in CSE-MRI. Two networks, respectively predicting water-dominant and fat-dominant tissue properties, produce two candidate solutions that can be compared based on likelihood to select the best tissue parameter estimate.

One may argue that the constraining transforms are not required because different networks may be learned by training multiple times until all possible variants are discovered. However, an obvious downside of this approach is an unnecessary increase in training time, which is not viable with this self-supervised implementation that requires to be retrained for different datasets.

Due to space limit, the present study has demonstrated our proposed method in only one qMRI example. However, model degeneracy is a common challenge in qMRI, from relaxometry~\cite{DeoniSeanC.L.2008GmTa} to diffusion~\cite{Jelescu_degen}. Evaluating the broad applicability of our approach will be a key focus of future work.

\clearpage

\begin{credits}
\subsubsection{\ackname} Giulio Minore is supported by the EPSRC funded UCL Centre for Doctoral Training in Intelligent, Integrated Imaging in Healthcare (i4health) (EP/S021930/1) and the National Institute for Health Research (NIHR) Biomedical Research Centre (BRC) at University College London Hospitals. Timothy Bray receives support from the NIHR BRC (personal support, and grants BRC1121/HEI/TB/110410 and BRC1185/III/TB/101350).


\subsubsection{\discintname}
The authors have no competing interests to declare that are
relevant to the content of this article.
\end{credits}
%
%
%
\bibliographystyle{splncs04}
\interlinepenalty=10000
\bibliography{mybibliography.bib}
%




\end{document}